\documentclass[twoside,fleqn]{article}
\usepackage{espcrc2}
\usepackage[dvips]{graphicx}
\newcommand{\be}{\begin{equation}}
\newcommand{\ee}{\end{equation}}

\title{
\vspace{-0.7cm}
\hfill \rm \null \hfill
 \hbox{\normalsize ADP-02-87/T526} \\
\vspace{-2mm}
 \hfill \hbox{\normalsize JLAB-THY-02-46} \\
\vspace{-3mm}
Spin-3/2 baryons in lattice QCD}

\author{%
J.~M.~Zanotti\address[CSSM]{Special Research Center for the
        Subatomic Structure of Matter, and
        Department of Physics and Mathematical Physics,
        University of Adelaide, 5005, Australia},
S.~Choe\address[HU]{Hiroshima Univ., Dept. of Physics, 
1-3-1 Kagamiyama, Higashi-Hiroshima 739-8526, Japan},
D.~B.~Leinweber\addressmark[CSSM],
W.~Melnitchouk\address[JLab]{Jefferson Lab, 12000
        Jefferson Avenue, Newport News, VA 23606, U.S.A.},
A.~G.~Williams\addressmark[CSSM]\thanks{Presented by A.~G.~Williams} 
and
J.~B.~Zhang\addressmark[CSSM]
}

\begin{document}

\thispagestyle{empty}

\begin{abstract}
  We present first results for masses of spin-${3/ 2}$ baryons in lattice
  QCD, using a novel fat-link clover fermion action in which only the
  irrelevant operators are constructed using fat links.  In the
  isospin-$1/ 2$ sector, we observe, after appropriate spin and parity
  projection, a strong signal for the $J^P={3\over 2}^-$ state, and find good
  agreement between the ${1\over 2}^+$ mass and earlier nucleon mass
  simulations with a spin-$1/ 2$ interpolating field.  For the
  isospin-$3/ 2$ $\Delta$ states, clear mass splittings are observed
  between the various ${1\over 2}^\pm$ and ${3\over 2}^\pm$ channels, with
  the calculated level orderings in good agreement with those observed
  empirically.
\end{abstract}

\maketitle





\section{INTRODUCTION}

Recent advances in computing power and the development of improved
actions have enabled simulations of the spectrum of excited states of
the nucleon in lattice QCD \cite{DEREK,DGR,other}.  The lattice
studies complement the high precision measurements of the $N^*$
spectrum under way at Jefferson Lab \cite{CLAS}.  In a recent paper
\cite{NSTAR}, results were presented for the excited nucleon and
spin-${1/ 2}$ hyperon spectra using the Fat-Link Irrelevant Clover
(FLIC) fermion action \cite{FATJAMES} with an ${\cal O}(a^2)$ improved
gluon action.  Here we extend the analysis of Ref.~\cite{NSTAR} to the
spin-${3/ 2}$ sector, and present results in both the isospin-${1/ 2}$
and ${3/ 2}$ channels.

In the isospin-${3/ 2}$ sector, our results for the $\Delta({3\over
  2}^+)$ state agree well with earlier simulations \cite{LDW} using
Wilson fermions.  We find a clear signal for the $P$-wave
$\Delta({3\over 2}^-)$ parity partner of the $\Delta$ ground state,
and a discernible signal for the $\Delta({1\over 2}^\pm)$ states.  In
particular, the $\Delta({1\over 2}^-)$ state is found to have a mass
$\sim 350$--400~MeV above the $\Delta({3\over 2}^+)$, with the
$\Delta({3\over 2}^-)$ slightly heavier.  The $\Delta({1\over 2}^+)$
state is found to lie $\sim 200$--300~MeV above these.  This level
ordering is consistent with experiment.

In the spin-${3/ 2}$ nucleon sector, there is good agreement for the
projected ${1\over 2}^+$ state with earlier nucleon mass calculations
\cite{NSTAR} using standard spin-${1/ 2}$ nucleon interpolating
fields.  Furthermore, we find a good signal for the $N({3\over 2}^-)$
state, with a mass splitting of $\sim 700$--1000~MeV with the nucleon
ground state.


\section{SPIN-${3\over 2}$ BARYONS ON THE LATTICE}

The standard isospin-${1/ 2}$, spin-${3/ 2}$, charge $+1$
interpolating field is given by
$\chi^{N^+}_{\mu}(x) = \epsilon^{abc}
\left( u^{Ta}(x)\ C \gamma_5 \gamma_\mu\ d^b(x) \right) \gamma_5 u^c(x)$,
which transforms as a pseudovector under parity, in accord with a
positive parity Rarita-Schwinger spinor.  The quark field operators
$u$ and $d$ act at Euclidean space-time point $x$, $C$ is the charge
conjugation matrix, $a, b$ and $c$ are color labels, and the
superscript $T$ denotes the transpose.  The charge neutral
interpolating field is obtained by interchanging $u \leftrightarrow d$.

The commonly used interpolating field for the $\Delta^{++}$ resonance
is given by \cite{IOFFE}
$\chi_\mu^{\Delta^{++}}(x) =
\epsilon^{abc} \left( u^{Ta}(x)\ C \gamma_\mu\ u^b(x) \right) u^c(x)$.
Since the spin-${3/ 2}$ Rarita-Schwinger spinor-vector is a tensor
product of a spin-1 vector and a spinor, the spin-${3/ 2}$ interpolating
field contains spin-${1/ 2}$ contributions.
To project a spin-${3/ 2}$ state one needs to use a spin-${3/ 2}$
projection operator \cite{BDM}.
Following spin projection, the correlation function for a given spin $s$,
$G^s_{\mu\nu}$, still contains positive and negative parity states.
In an analogous procedure to that used in Ref.~\cite{NSTAR}, where a
fixed boundary condition is used in the time direction, positive and
negative parity states are obtained by taking the trace of
$G^s_{\mu\nu}$ with the operators
$\Gamma_{\pm} = {1\over 2}\left( 1\pm \gamma_4 \right)$, respectively.


\section{RESULTS}

Lattice simulations are performed on a $16^3\times 32$ lattice at
$\beta=4.60$, corresponding to a lattice spacing of $a = 0.122(2)$~fm
set by the string tension \cite{Edwards:1997xf} with $\sqrt\sigma =
440$~MeV.
We consider 392 configurations generated on the Orion cluster at the
CSSM, U. Adelaide.
A mean-field improved plaquette plus rectangle gauge action is used.
For the quark fields, a Fat-Link Irrelevant Clover (FLIC)
\cite{FATJAMES} action is implemented.
We employ a highly improved definition of $F_{\mu\nu}$
\cite{FATJAMES,SUNDANCE} leaving errors of ${\cal O}(a^6)$.
Mean-field improvement of the tree-level clover coefficient with fat
links represents a small correction and proves to be quite adequate
\cite{FATJAMES}.
A fixed boundary condition in the time direction is used for the
fermions by setting $U_t(\vec x, N_t) = 0\ \forall\ \vec x$ in the
hopping terms of the fermion action, with periodic boundary conditions
imposed in the spatial directions.
Gauge-invariant gaussian smearing in the spatial dimensions is applied
at the fermion source (placed at time slice $t=3$) to increase the
overlap of the interpolating operators with the ground states.

%
In Fig.~1 the results for the spin-projected $\Delta({3\over 2}^+)$
(triangles) and $\Delta({3\over 2}^-)$ (diamonds) masses are shown
as a function of the pseudoscalar meson mass, $m_\pi^2$.
The trend of the $\Delta({3\over 2}^+)$ data points with decreasing
$m_q$ is consistent with the physical mass of the $\Delta(1232)$,
with some nonlinearity in $m_\pi^2$ expected near the chiral limit.
The mass of the $\Delta({3\over 2}^-)$ lies some 500~MeV above that
of its parity partner.

When performing a spin projection to extract the
$\Delta({1\over 2}^\pm)$ states, a discernible, although noisy, signal
is detected.
The $\chi_\mu^{\Delta^{++}}$ interpolating field has small overlap
with spin-${1/ 2}$ states, however, with $\sim 400$ configurations
we are able to extract masses for the spin-${1/ 2}$ states at
sufficiently early times.
The $\Delta({1 \over 2}^+)$ (squares) and $\Delta({1 \over 2}^-)$
(open circles) are also displayed in Fig.~1.
The lowest excitation of the ground state, namely the
$\Delta({1\over 2}^-)$, has a mass $\sim 350$--400~MeV above the
$\Delta({3\over 2}^+)$, with the $\Delta({3\over 2}^-)$ slightly
heavier.
The $\Delta({1\over 2}^+)$ state is found to lie $\sim 200$--300~MeV
above these, although the signal becomes weak at smaller quark masses.
This level ordering is consistent with that observed in the empirical
mass spectrum.

\begin{figure}[t]
\begin{center}
{\includegraphics[height=7cm,angle=90]{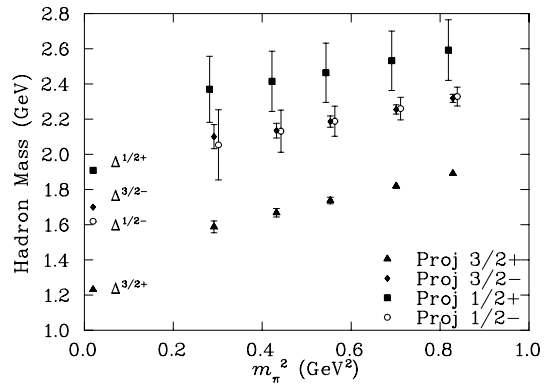}}
\vspace*{-0.5cm}
\caption{Masses of the spin-projected $\Delta({3\over 2}^\pm)$
  and $\Delta({1\over 2}^\pm)$ states.
  The empirical masses are indicated along the ordinate. \vspace*{-9mm}}
\end{center}
\end{figure}

In the isospin-${1/ 2}$ sector, large statistical fluctuations make it
difficult to obtain a clear signal, even with 392 configurations.  A
reasonable value of $\chi^2 / N_{\rm DF}$ is obtained for time slices
$t=7$--8.
These results are shown as a function of $m_{\pi}^2$ in Fig.~2.
Parity projecting to extract the $N({3\over 2}^+)$ state, we find
that the correlation function changes sign and has a large negative
contribution in the range $t=7$--11.
This behavior is an artifact associated with the quenched decay of
the excited state into $N + \eta'$, and is further explored in
Ref.~\cite{fullSpin32}.

The $N({1\over 2}^+)$ channel displays the interplay of a quenched decay
channel and the ground state contribution.
A strong $P$-wave coupling of the $N({1\over 2}^+)$ to $N \eta'$ forces
the correlation function to be negative at small times, which then turns
positive at larger times when the ground state contribution begins to
dominate the correlation function.
A good $\chi^2 / N_{\rm DF}$ value is obtained for a fit to time slices
$t=10$--14.
This suggests that the first excited state has strong coupling to the
$\eta'$, which implies a gluon-rich structure for the Roper
resonance~\cite{ROPER}.

\begin{figure}[t]
\begin{center}
{\includegraphics[height=7cm,angle=90]{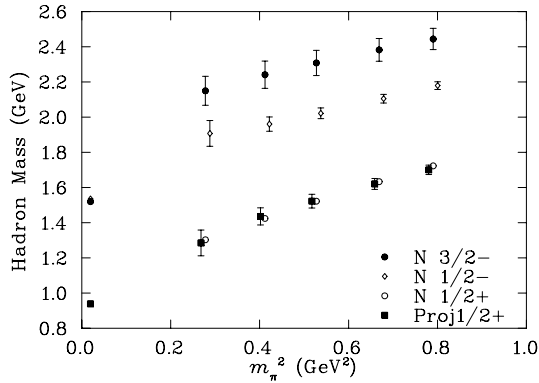}}
\vspace*{-0.5cm}
\caption{Masses of the spin-projected $N({3\over 2}^-)$ and
        $N({1\over 2}^+)$ states, compared with
        the nucleon and $N({1\over 2}^-)$ masses from
        Ref.~\protect\cite{NSTAR}.\vspace*{-9mm}}
\end{center}
\end{figure}

The extracted masses of the $N({3\over 2}^-)$ and $N({1\over 2}^+)$
states are displayed in Fig.~2 as a function of $m_\pi^2$.
Earlier results using the standard spin-${1/ 2}$ interpolating
field \cite{NSTAR,FATJAMES} are also shown in Fig.~2 for reference.
There is excellent agreement between the spin-projected ${1\over 2}^+$
state obtained from the spin-${3/ 2}$ interpolating field and
the earlier ${1\over 2}^+$ results.
A clear mass splitting is also seen between the $N({3\over 2}^-)$
and $N({1\over 2}^+)$ states obtained from the spin-${3/ 2}$
interpolating field, with a mass difference of 700--1000~MeV.

\section{CONCLUSION}

First results for the spectrum of spin-${3/ 2}$ baryons in the
isospin-${1/ 2}$ and ${3/ 2}$ channels are reported, using a
FLIC fermion action and an ${\cal O}(a^2)$ mean-field improved gauge
action.
Good agreement is found with earlier calculations for the $\Delta$
ground state, and clear mass splittings between the ground state and
its parity partner are observed.
A signal is also obtained for the $\Delta({1\over 2}^\pm)$ states,
with the level ordering consistent with the observed empirical
mass spectrum.
 
For isospin-${1/ 2}$ baryons, clear signals are obtained for
both the $N({3\over 2}^-)$ and spin-projected $N({1\over 2}^+)$ states
from a spin-${3/ 2}$ interpolating field.
The ${1\over 2}^+$ state in particular is in good agreement with earlier
simulations of the nucleon mass using standard spin-${1/ 2}$
interpolating fields.

\vspace*{0.3cm}

We thank R.G.~Edwards and D.G.~Richards for helpful discussions.
This work was supported by the Australian Research Council.
W.M. is supported by the U.S. Department of Energy contract
\mbox{DE-AC05-84ER40150}
under which the Southeastern Universities Research Association
(SURA) operates the Thomas Jefferson National Accelerator Facility
(Jefferson Lab).



\end{document}